\documentstyle[aps,prd]{revtex}

\begin{document}

\def\be{\begin{equation}}
\def\ee{\end{equation}}
\def\beq{\begin{eqnarray}}
\def\eeq{\end{eqnarray}}
\def\s{\sigma}
\def\p{\rho}
\def\G{\Gamma}
\def\F{_2F_1}
\def\an{analytic}
\def\ac{\an{} continuation}
\def\hsr{hypergeometric series representations}
\def\hf{hypergeometric function}
\def\ndim{NDIM}
\def\quarto{\frac{1}{4}}
\def\half{\frac{1}{2}}

\draft
\title{Negative dimensional approach for scalar two-loop three-point
and three-loop two-point integrals}
\author{A. T. Suzuki and A. G. M. Schmidt}
\address{Universidade Estadual Paulista -- Instituto de F\'{\i}sica 
Te\'orica, R.Pamplona, 145, S\~ao Paulo SP, CEP 01405-900, Brazil }
\date{\today}

\maketitle

\begin{abstract}
The well-known $D$-dimensional Feynman integrals were shown, by Halliday and
Ricotta, to be capable of undergoing \ac{} into the domain of negative values
for the dimension of space-time. Furthermore, this could be identified with
Grassmannian integration in positive dimensions. From this possibility follows
the concept of negative dimensional integration for loop integrals in field
theories. Using this technique, we evaluate three two-loop three-point scalar
integrals, with five and six massless propagators, with specific external
kinematic configurations (two legs on-shell), and four three-loop
two-point scalar integrals. These results are given for arbitrary exponents of
propagators and dimension, in Euclidean space, and the particular cases
compared to results published in the literature.
\end{abstract}


\section{Introduction.}

The study of quantum field theories in the perturbative regime can sometimes
become a very challenging and arduous issue to deal with, especially when one
needs to go on calculating multi-loop Feynman integrals. There are several
techniques that have been developed and refined in the course of time, but
usually the calculations are rather involved and analytic results are difficult
to obtain. Depending on the approach one uses to tackle the problem, the
obstacles to surmount can be more or less demanding. For example, if one uses
the standard Feynman parametrization, the integral over momentum space is
straightforward --- in the sense that nowadays one can easily fint them out even
tabulated in textbooks on quantum field theories --- but the resulting
parametric integrals that are left over become quite complicated and cumbersome
to tackle, especially with increasing number of internal momenta in the loops.
Of course, if one is clever enough, he/she may be able to overcome some of
these hurdles and even perform really hard calculations
\cite{kramer,gonsalves}.

There are, of course, other techniques which have been considered in the
literature, e.g., integration by parts and Gegenbauer polynomial method in
configuration space \cite{russo}; the Mellin-Barnes representation for massive
propagators \cite{boos} to cite just a few of them \cite{davyd}, each one of
those with its own strengths and weaknesses.

On the other hand, there is this novel technique known as negative dimensional
integration method (\ndim{}), first devised by Halliday and Ricotta
\cite{halliday}. It has as its starting point the principle of \ac{}, it has
the remarkable property of being equivalent to Grassmannian integration in
(positive) $D$-dimensional space \cite{halliday2}, it requires only the
integration of Gaussian-like integrals and solving of linear algebraic
equations and is suitable to deal with massless \cite{lab} or massive
\cite{box} diagrams in on- and off-shell regimes \cite{preprint} and even to
carry out light-cone gauge integrals, with added troublesome gauge-dependent
poles \cite{probing}. 

Our aim in this work is to demonstrate the simplicity of NDIM calculating some
scalar three-point integrals, with five and six massless propagators, at
two-loop level and some two-point three-loop integrals also in the massless
case. The outline for our work is as follows: in section II we present a
detailed calculation of a two-loop integral in the \ndim{} approach; in section
III we write down the results for the remaining two- and three-loop scalar
integrals, while in section IV we give our concluding remarks about this work.

\section{Two-loop three-point vertex with six massless propagators.}

This computation is performed following just a few simple steps
\cite{lab}.
Fist of all, let us consider the Gaussian-like integral,

\beq 
\label{gauss} 
F &=& \int\int d^D\!r\; d^D\!q \;\;\exp\left[-\alpha q^2 -\beta r^2
-\gamma (q-r)^2 -\delta (p+q)^2 -\omega(q-k)^2-\theta (q-r-k)^2\right]\nonumber\\
&=&  \left(\frac{\pi^2}{\lambda}\right)^{D/2} \exp{\left[ -
\frac{1}{\lambda}\left(\alpha\beta\delta+ \alpha\gamma\delta+
\beta\gamma\delta+\alpha\theta\delta\right)p^2\right]} ,
\eeq
where the arguments in the exponential function of the integrand correspond to
propagators in the diagram of Fig.1, with $k^2=t^2=0$ satisfying the on-shell
condition. For the sake of compactness, we define
\[
\lambda=\alpha\beta+\alpha\gamma+\alpha\theta+\beta\gamma+\beta\delta+
\gamma\delta+\theta\delta+\beta\omega+\gamma\omega+\theta\omega+\beta\theta.
\]
The parameters $(\alpha,\beta,\gamma,\delta,\omega,\theta)$ are quite general
and arbitrary except that their real parts must be positive to make sure we
have well-defined objects over the whole integration region.

Expanding the exponential in the second line of (\ref{gauss}) in Taylor series
and using the multinomial expansion for the argument of the exponential, we
obtain,

\beq 
F &=& \pi^D\sum_{\{x,y,z=0\}}^\infty \frac{(-p^2)^{\Sigma x} (-\Sigma
x-D/2)! \alpha^a \beta^b \gamma^c \delta^d \omega^e \theta^f}{x_1!\cdot\cdot
\cdot x_4!y_1!\cdot\cdot\cdot y_9!z_1!z_2!} \delta_{-\Sigma x-D/2,\Sigma
y+\Sigma z},
\eeq

where we have a fifteen-fold sum, with $\Sigma x=x_1+x_2+x_3+x_4$ and we
define,

\beq
a&=&x_1+x_2+x_4+y_1+y_2+y_3\:,\nonumber\\
b&=&x_1+x_3+y_1+y_4+y_5+y_8+z_2\:,\nonumber\\
c&=&x_2+x_3+y_2+y_4+y_6+y_9\:,\nonumber\\
d&=&x_1+x_2+x_3+x_4+y_5+y_6+y_7\:,\nonumber\\
e&=&y_8+y_9+z_1\:,\nonumber\\
f&=&x_4+y_3+y_7+z_1+z_2\:.
\eeq

On the other hand, taking the Taylor expansion of the exponential that is under
the integration sign in (\ref{gauss}), we have,

\begin{equation} 
\label{proj} F = \sum_{i,j,l,m,n,s=0}^\infty \frac{(-1)^{i+j+l+m+n+s}}{i!j!l!m!n!s!}
\alpha^i\beta^j\gamma^l\delta^m\omega^n\theta^s {\cal S}_{NDIM}^1, \ee{}
where ${\cal {S}}_{NDIM}^1$ is the relevant integral in negative $D$, defined
by 
 \be \label{Indim} {\cal S}_{NDIM}^1 = \int d^D\!q\;d^D\!r
(q^2)^i(r^2)^j(q-r)^{2l}(q+p)^{2m} (q-k)^{2n}(q-r-k)^{2s} .
\ee

Now comparing both expressions for the original integral $F$ we are led to the
conclusion that, in order to have equality between these two expressions, the
factor ${\cal S}_{NDIM}^1$ must be given by the multiple series,

\be
\label{geral} 
{\cal S}_{NDIM}^1 = P_1 \sum_{x,y,z=0}^\infty
\frac{1}{x_1!\cdot\cdot\cdot x_4!y_1!\cdot\cdot\cdot y_9!z_1!z_2!}
 \; \delta_{a,i}\;\delta_{b,j}\;\delta_{c,l}\;\delta_{d,m}\;\delta_{e,n}
 \;\delta_{f,s} ,
\ee
with 
\[ P_1 = (-\pi)^D(p^2)^\s\G(1+i)\G(1+j)\G(1+l)\G(1+m)\G(1+n)\G(1+s)
\G(1-\s-\half D),
\]
and for convenience we use the definition $\s=i+j+l+m+n+s+D$. The system we
must solve has, therefore, fifteen ``unknowns'' (the indices $x,y,z$) and only
seven equations --- six of which comes from comparing the exponents of
$(\alpha,\beta,\cdots)$ and one from the multinomial expansion, namely, $\Sigma
y+\Sigma z=-\Sigma x-D/2$. So, such a system can only be solved in terms of
some of these ``unknowns'', and we conclude, from the combinatorics, that there
are $C^{15}_7=6435$ systems $(7\times 7)$ of linear algebraic equations, i.e.,
there are $6435$ different ways of choosing these ``unknowns''. Of this grand
total possible solutions, $4113$ yield trivial solutions and for this reason
present no interest at all. Yet, the remaining $2322$ solutions are quite a big
number! Fortunately, not all the remnant non-vanishing solutions need to be
considered in order to solve the Feynman integral. In fact, only linearly
independent series will be needed. In our massless case and where there is only
one non-vanishing external momentum, the argument of such series is unit. Since
NDIM provides, in general, hypergeometric series (see, for example ref.
\cite{probing} where we have double and triple series of such), we obtain in
this case $_pF_q(\cdots|1)$, and from the theory of generalized hypergeometric
functions \cite{bailey,luke}, a linear combination of linearly independent
series of this type has up to $p$ series.

Consider then, the solution where the indices $y_2, y_3, y_4, y_5, y_7, y_8, y_9,
z_2$ are left undetermined. Solving this system we get,
  
\beq
x_1&=&-l+m+n-s+y_2+y_3+y_4-y_5-y_8+z_2\:,\nonumber\\
x_2&=&i+l+n+D/2-y_2-y_9+y_5+y_7+z_2\:,\nonumber\\
x_3&=&j+l+s+D/2-y_4-z_2\:,\nonumber\\
x_4&=&-n+s-y_3-y_7+y_8+y_9-z_2\:,\nonumber\\
y_1&=&-m-n-D/2-y_2-y_3-y_4-z_2\:,\nonumber\\
y_6&=&-\s+m-y_5-y_7\:,\nonumber\\
z_1&=&n-y_8-y_9\:, 
\eeq
which substituted in (\ref{geral}) yields an eight-fold summation. Seven of
these can be rewritten in terms of gamma functions. The procedure is as
follows: Firstly, the factorials must be converted into gamma functions and
these into Pochhammer symbols, defined by,

\be 
(n|k) \equiv  (n)_k = \frac{\G(n+k)}{\G(n)} .
\ee
Secondly, use the properties,

\be
\label{prop} 
(n|j+k) = (n+j|k)(n|j),\qquad\qquad
(n|-k) = \frac{(-1)^k}{(1-n|k)}, 
\ee
which follow from the definition above. Then, it is possible to rewrite some
series in terms of Gaussian \hf{}s, which can, in general, be summed when its
argument is unit \cite{bateman}. 

\be
\label{somaf21} 
_2F_1(a,b;c|1) = \frac{\G(c)\G(c-a-b)}{\G(c-a)\G(c-b)}.
\ee
Exception for this occurs when the parameters are such that $c-a-b=0$. 

Therefore, if we rearrange these series conveniently, we can sum them, and
applying this procedure for the series defined by indices $y_4, y_3, y_2, y_8,
y_9, y_7, y_5$, they all can be summed out. The last series, defined by index
$z_2$, however cannot be summed in this fashion since it is a $_3F_2$ function,
which is summable only when it is of Saalsch\"utzian type \cite{bailey}, and
this is not our present case.

Many of the gamma functions that appear in this process simplify among
themselves and in the end, what remains are fourteen gamma functions
distributed in a fraction, i.e., seven in the numerator and seven in the
denominator multiplying the $z_2$ series. This is the result in the {\it
negative} dimension region and positive exponents $(i,j,l,m,n,s)$, see
(\ref{proj}). Now we must be able to bring this result back, that is,
analytically continue it, to our real physical world, that of positive $D$. This
is carried out very easily: group the gamma functions in convenient Pochhammer
symbols and use the second property (\ref{prop}).

Considering now the solution where $x_1, x_2, x_3, y_1, y_5, y_6, y_9, z_2$ are
left undetermined, we get another $_3F_2(a,b,c;e,f|z)$ function. Let us recall
that such functions have three singular points, namely $(0,1,\infty)$. Here, we
are interested in solutions in the vicinity of $z=1$. No$\!\!\!$/rlund \cite{norlund}
proved that when the difference between $e$ and $f$ is an integer, some of the
series concide or are without meaning. This is precisely our case. This means
that our final result for the Feynman integral shall exhibit only two linearly
independent generalized \hf{}s of the type $_3F_2$, instead of the usual three.
We get, as a result, then

\be
{\cal S}_{NDIM}^1 = \pi^D (p^2)^\s \{{\bf A}\:_3F_2(a_1,b_1,c_1;e_1,f_1|1)+{\bf
B}\:_3F_2(a'_1,b'_1,c'_1;e'_1,f'_1|1)\}
\ee
where
\beq
{\bf A}&\equiv & (-m|\s)(\s+D/2|-2\s-D/2) (-j|j+l+s+D/2)(-l|j+l+s+D/2)
\nonumber\\
&\times &(-i-j-l-s-D/2|i)(j+l+s+D|-j-l-s-D/2+m+n),
\eeq
and
\beq
{\bf B}&\equiv & (-i|\s)(-m|\s)(\s+D/2|-2\s-D/2)\nonumber\\
&\times &(-j|-l-s-D/2)(j+l+s+D|-j-D/2) (-l-s|j+l+s+D/2),
\eeq
where the set of parameters are defined as
$(a_1,\,b_1,\,c_1;\,e_1,\,f_1)=(-s,\,m+n+D/2,\,-j-l-s-D/2;\,-i-j-l-s-D/2,\,1-j-s-D/2)$,
and
$(a'_1,\,b'_1,\,c'_1;\,e'_1,\,f'_1)=(-\s,\,-l,\,-j-l-s-D/2;\,-l-s,\,1+i-\s)$.
Note that $(e_1-f_1)$ is an integer. Observe also that, in the special case
where all the exponents are equal to minus one, the hypergeometric function
$_3F_2$ reduces to $_2F_1$, which can be summed again, using (\ref{somaf21}).
So, for this special case of negative unity exponents, the final result is
expressed in terms of gamma functions only.

Let us point out here that our final result agrees with that given by
Davydychev and Osland \cite{davyd}, i.e., meaning that equation (27) in ref.
\cite{kramer} disagrees with ours. The difference is in the $9\zeta(4)$ factor
in \cite{kramer} which should correctly be read $9\zeta(2)$. This gives the
correct result for $i=j=l=m=n=s=-1$ and $D=4-2\epsilon$.

\section{Results for some two and three-loop massless scalar integrals.}

The scalar integrals for the remaining diagrams (Fig.2-7) are, in negative $D$,
given by (number superscripts 2-7 label the corresponding diagrams):

\be {\cal S}_{NDIM}^2 = \int d^D\!q\;d^D\!r \; (r^2)^i (q^2)^j
(p+r)^{2l}(q+k)^{2m} (p+r-q)^{2n}, \ee

\be {\cal S}_{NDIM}^3 = \int d^D\!q\;d^D\!r \; (r^2)^i (q^2)^j
(k-q-r)^{2l}(k-r)^{2m} (r+p)^{2n}, \ee

\be {\cal S}_{NDIM}^4 = \int d^D\!q\;d^D\!r\;d^D\!k \; (r^2)^i (k^2)^j
(q^2)^l (p-q-r)^{2m} (r-k)^{2n}, \ee

\be {\cal S}_{NDIM}^5 = \int d^D\!q\;d^D\!r\;d^D\!k \; (q^2)^i (k^2)^j
(r^2)^l (p-r)^{2m} (r-q)^{2n} (p-r-k)^{2s}, \ee

\be {\cal S}_{NDIM}^6 = \int d^D\!q\;d^D\!r\;d^D\!k \; (r^2)^i (k^2)^j
(q^2)^l (r+k)^{2m}(p-k-r)^{2n} (r+k-q)^{2s}, \ee
and 
\be {\cal S}_{NDIM}^7 = \int d^D\!k\;d^D\!q\;d^D\!r\;
(q^2)^i(k^2)^j (p-k-r)^{2l} (k-q)^{2m} (p-k)^{2n} (r^2)^s .\ee

All these can be evaluated in a similar way following the steps of the previous
section. For this reason we quote here only the final results, after
continuation to positive dimension and negative exponents of propagators:

\be
{\cal S}_{NDIM}^2 = \pi^D (k^2)^\p \{{\bf C}\:_3F_2(a_2,b_2,c_2;e_2,f_2|1)+{\bf
D}\:_3F_2(a'_2,b'_2,c'_2;e'_2,f'_2|1)\} ,
\ee
where
\beq
{\bf C}&\equiv & (-i|\p)(\p+D/2|-2\p-D/2)(i+D/2-\p|\p)(\p-i-l|-j-D/2) \nonumber\\
&\times &(-j|-m-n-D/2)(-n|j+m+n+D/2)\;
\eeq
and
\beq
{\bf D}&\equiv & (-i|\p)(-l|\p)(\p+D/2|-2\p-D/2)(j+m+n+D|-j-m-D/2)\nonumber\\
&\times &(-j-m|-n-D/2)(-n|j+m+n+D/2)\; ,
\eeq
with the set of parameters defined by
$(a_2,\,b_2,\,c_2;\,e_2,\,f_2)=(-m,\,i+D/2,\,-j-m-n-D/2;\,i+D/2-\p,\,1-m-n-D/2)$
and $(a'_2,\,b'_2,\,c'_2;\,e'_2,\,f'_2)=(-j,\,-\p,\,-j-m-n-D/2;1+l-\p,\,-j-m)$
and $\p=\s-s$. Note that here also we have a linear combination of two $_3F_2$
\hf{}s, since $(e_2-f_2)$ is an integer and No$\!\!\!/$rlund's theorem applies.

\beq 
{\cal S}_{NDIM}^3 &=& \pi^D (k^2)^\p
(-l|\p)(\p+D/2|-2\p-D/2)(i+l+D/2-\p|\p)(-j|j+n+D/2) \nonumber\\
&&\times (-n|j+n+D/2)(j+n+D|-2j-2n-3D/2),
\eeq

\beq {\cal S}_{NDIM}^4 &=& \pi^{3D/2}(p^2)^{\p'} (-j|j+n+D/2)(-n|j+n+D/2)
(-l|l+m+D/2)\nonumber\\
&&\times (-m|l+m+D/2) (\p'+D/2|-2\p'-D/2)(j+n+D|i)\nonumber\\
&&\times (-j-n-D/2|-i),
\eeq
where $\p'=\p+D/2$.

\beq 
{\cal S}_{NDIM}^5 &=& \pi^{3D/2}(p^2)^{\s'} (-i|i+n+D/2)(-n|i+n+D/2)
(-j|j+s+D/2)\nonumber\\
&& \times (-s|j+s+D/2)(-i-l-n-D/2|l)(-j-m-s-D/2|m)\nonumber\\
&&\times (\s'+D/2|-2\s'-D/2)(i+n+D|l)(j+s+D|m),
\eeq
where $\s'=\s+D/2$, 

\beq 
{\cal S}_{NDIM}^6 &=& \pi^{3D/2}(p^2)^{\s'} (-i|i+j+D/2)(-j|i+j+D/2)
(-l|l+s+D/2)(-s|l+s+D/2) \nonumber\\
&& \times (\s'+D/2|-2\s'-D/2)(-i-j-l-m-s-D|l+m+s+D/2)
\nonumber\\
&&\times (i+j+D|l+m+s+D/2)(-n|-l-s-D/2+n)\nonumber\\
&&\times (l+s+D|-l-s-D/2+n). 
\eeq

Note that this result contains a special case (for $m=0$) calculated
numerically in \cite{easther}. Our result, on the other hand, was analytically
obtained and is more general, since the exponents of propagators are arbitrary
(cf. results in the second reference in \cite{lab}).

Finally,
\beq 
{\cal S}_{NDIM}^7 &=& \pi^{3D/2} (p^2)^{\s'}
(-i|i+m+D/2)(-m|i+m+D/2)(-l|l+s+D/2)\\
 &&\times
(\s'+D/2|-2\s'-D/2)(-i-j-m-D/2|j)(l+s+D|-2l-2s-3D/2)
 \nonumber\\ 
&&\times(-l-n-s-D/2|\s')(i+m+D|-i+l-m+n+s)(-s|l+s+D/2)\nonumber .
\eeq

\section{Conclusion.}

Although \ndim{} is not a regularization method, it shares all the concepts of
dimensional regularization since it is based on the principle of analytic
continuation in the dimension of space-time. In this work we showed that with
an adequate technique such as \ndim{} it is possible to calculate some two- and
three-loop Feynman integrals with relative easiness and without much elaborate
machinery, just the performance of some Gaussian-type integrals plus resolution
of systems of linear algebraic equations. Of course, in view of the great
number of systems of linear algebraic equations that are involved, we need the
help of computer facilities, but the task is simple, which a PC can handle
properly. 

The results we obtained can be used to study on-shell form factors in massless
QCD at the two-loop level. Moreover, to illustrate how \ndim{} works beyond
two-loops, we also performed some calculations of scalar three-loop, two-point
integrals with arbitrary exponents for propagators and dimension, in Euclidean
space.

\acknowledgments{A.G.M.S. gratefully acknowledges FAPESP (Funda\c c\~ao de Amparo \`a Pesquisa do
Estado de S\~ao Paulo, Brasil) for financial support.}

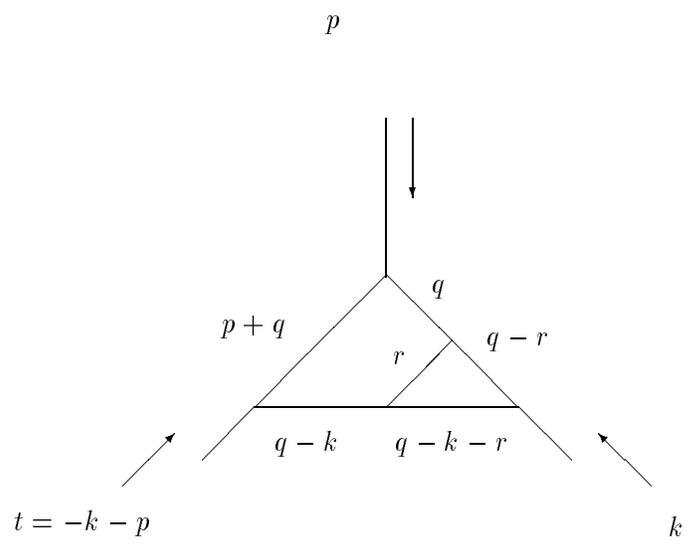
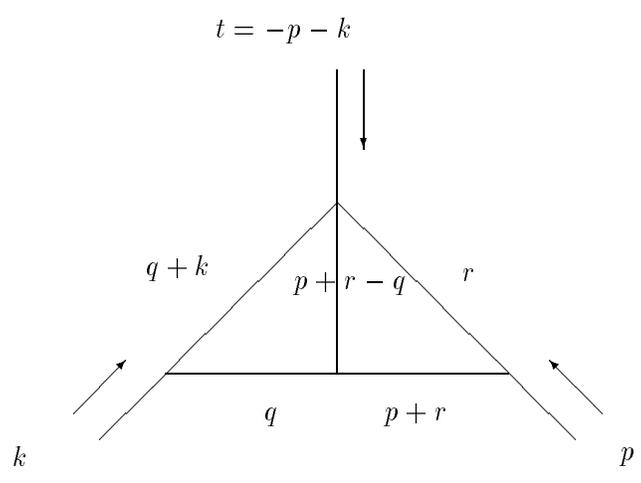
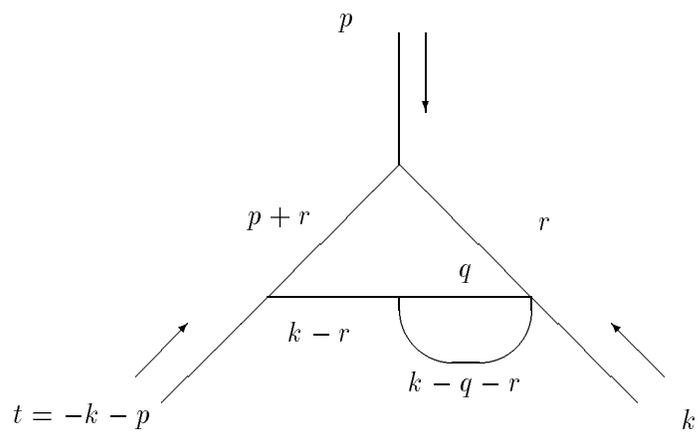
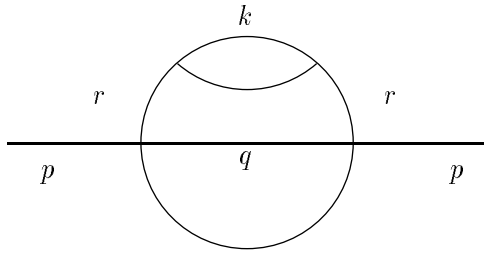
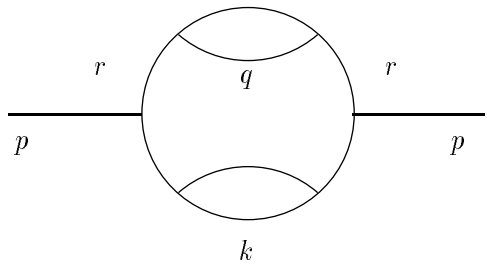
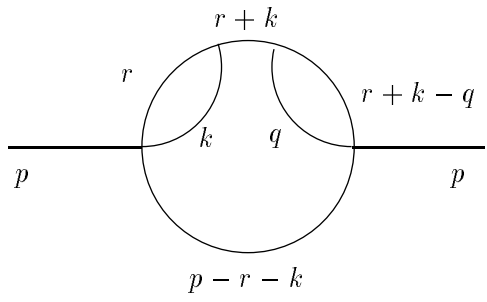
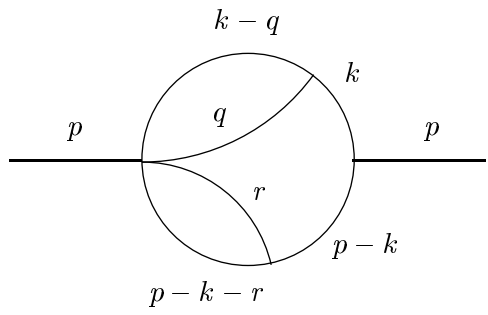
\end{document}